\documentclass[12pt,preprint]{aastex}

\newcommand{\ldl}{$\lambda/{\Delta}{\lambda}$}
\newcommand{\teff}{T$_{eff}$}
\newcommand{\logg}{$\log{g}$}

\newcommand{\meth}{CH$_4$}
\newcommand{\water}{H$_2$O}
\newcommand{\namesrc}{HD~3651}

\slugcomment{Submitted to ApJ 6 Sep 2006; accepted 16 Nov 2006}

\shorttitle{Physical Properties of HD~3651B}
\shortauthors{Burgasser et al.}

\begin{document}

\title{The Physical Properties of HD~3651B: An Extrasolar Nemesis?}

\author{Adam J.\ Burgasser\altaffilmark{1}}

\affil{Massachusetts Institute of Technology, Kavli Institute for Astrophysics and Space Research,
Building 37, Room 664B, 77 Massachusetts Avenue, Cambridge, MA 02139; ajb@mit.edu}

\altaffiltext{1}{Visiting Astronomer at the Infrared Telescope Facility, which is operated by
the University of Hawaii under Cooperative Agreement NCC 5-538 with the National Aeronautics
and Space Administration, Office of Space Science, Planetary Astronomy Program.}

\begin{abstract}
I present detailed analysis of the  
near infrared spectrum of HD~3651B, a faint, co-moving wide companion 
to the nearby planet-hosting star HD~3651.  These data confirm the companion as a brown dwarf with spectral type T8, consistent with the analysis of Luhman et al.  Application of the semi-empirical technique of Burgasser,
Burrows \& Kirkpatrick indicates that HD~3651B has $T_{eff}$ = 790$\pm$30~K and $\log{g}$ = 5.0$\pm$0.3 for a metallicity of [M/H] = 0.12$\pm$0.04, 
consistent with a
mass M = 0.033$\pm$0.013~M$_{\sun}$ and an age of 0.7--4.7~Gyr.
The surface gravity, mass and age estimates of this source 
are all highly sensitive to the 
assumed metallicity; however, a supersolar metallicity is
deduced by direct comparison of spectral models to 
the observed absolute fluxes.  The age of HD~3651B is somewhat better
constrained than that of the primary, with estimates for the latter ranging
over $\sim$2~Gyr to $>$12~Gyr.
As a widely orbiting massive object to a known planetary
system that could potentially 
harbor terrestrial planets in its habitable zone, HD~3651B
may play the role of Nemesis in this system.
\end{abstract}

\keywords{Galaxy: solar neighborhood ---
stars: individual (HD~3651B, Gliese~570D) ---
stars: fundamental parameters ---
stars: low mass, brown dwarfs ---
stars: planetary systems}

\section{Introduction}

\citet{luh06} and \citet{mug06} have recently reported the discovery of a co-moving,
widely-separated
(43$\arcsec$; 480~AU projected separation) companion to the nearby K0~V dwarf
{\namesrc}.  Low resolution near-infrared spectroscopy of this source
has identified it as a late T-type brown dwarf \citep{luh06}.
The {\namesrc} system also
hosts a closely-separated ($a$ = 0.296$\pm$0.017~AU), high-eccentricity
($e$ = 0.64$\pm$0.04) sub-Saturn-mass planet
(M$\sin{i}$ = 0.23$\pm$0.02 M$_{Jupiter}$),
{\namesrc}b, identified
from radial velocity variability \citep{fis03}.  While several
exoplanet-host stars are known to have
stellar-mass companions (e.g., \citealt{low02,pat02,egg04,rag06}),
{\namesrc}ABb is the
first star/brown dwarf/planet system to be identified, and is
therefore a key target for
studies of all three of these mass-delineated classes.

In this article, I analyze low-resolution near-infrared spectroscopy 
of {\namesrc}B, using the technique of \citet[hereafter BBK]{metgrav}
to derive the effective temperature ({\teff}), surface gravity
({\logg}), metallicity ([M/H]), mass and age of this source.  
Acquisition and characterization 
of the spectral data, obtained with SpeX 
\citep{ray03} mounted on the 3m
NASA Infrared Telescope Facility (hereafter IRTF) is described in $\S$~2. Spectral analysis to derive the physical properties of HD~3651B is presented in  $\S$~3, with the importance of metallicity effects highlighted by the
determination that this companion must be 
metal-rich like the system's primary.
Finally, in $\S$~5 I
speculate on the role of {\namesrc}B as a Nemesis analog 
in the {\namesrc} system.

\section{Observations}

{\namesrc}B was observed with SpeX on 3 September 2006 (UT) under clear
and dry conditions with good seeing (0$\farcs$7 at $J$-band). 
The target was acquired using the SpeX guiding camera, and 
low resolution near infrared
spectral data were obtained
using the SpeX prism mode and 0$\farcs$5 slit, yielding
a spectral resolution {\ldl} $\approx$ 120.
%and dispersion across the chip of 20--30~{\AA}~pixel$^{-1}$.
The slit was aligned to the parallactic angle to mitigate 
differential refraction effects. 
Note that this setup was similar to that used by 
\citet{luh06} for their spectral
observations of {\namesrc}B, with the exception
of a slightly narrower slit (0$\farcs$5 versus 0$\farcs$8).
Eight exposures of 180~s each
were obtained in an ABBA dither pattern along the slit.
%, for a total
%integration time of 1440~s at airmass 1.06.  
%As {\namesrc}B is quite
%faint, the telescope was guided on a nearby source present in the
%guide camera's field of view.  
The A0~V star HD~7215 was observed immediately
afterward and at a similar airmass (1.06) for flux calibration.
Internal flat field and Ar arc lamps exposures were also acquired
for pixel response and wavelength calibration.  All data were
reduced using SpeXtool version 3.3 
\citep{cus04} using standard settings (cf.~\citealt{mewide3}).

\begin{figure}
\plotone{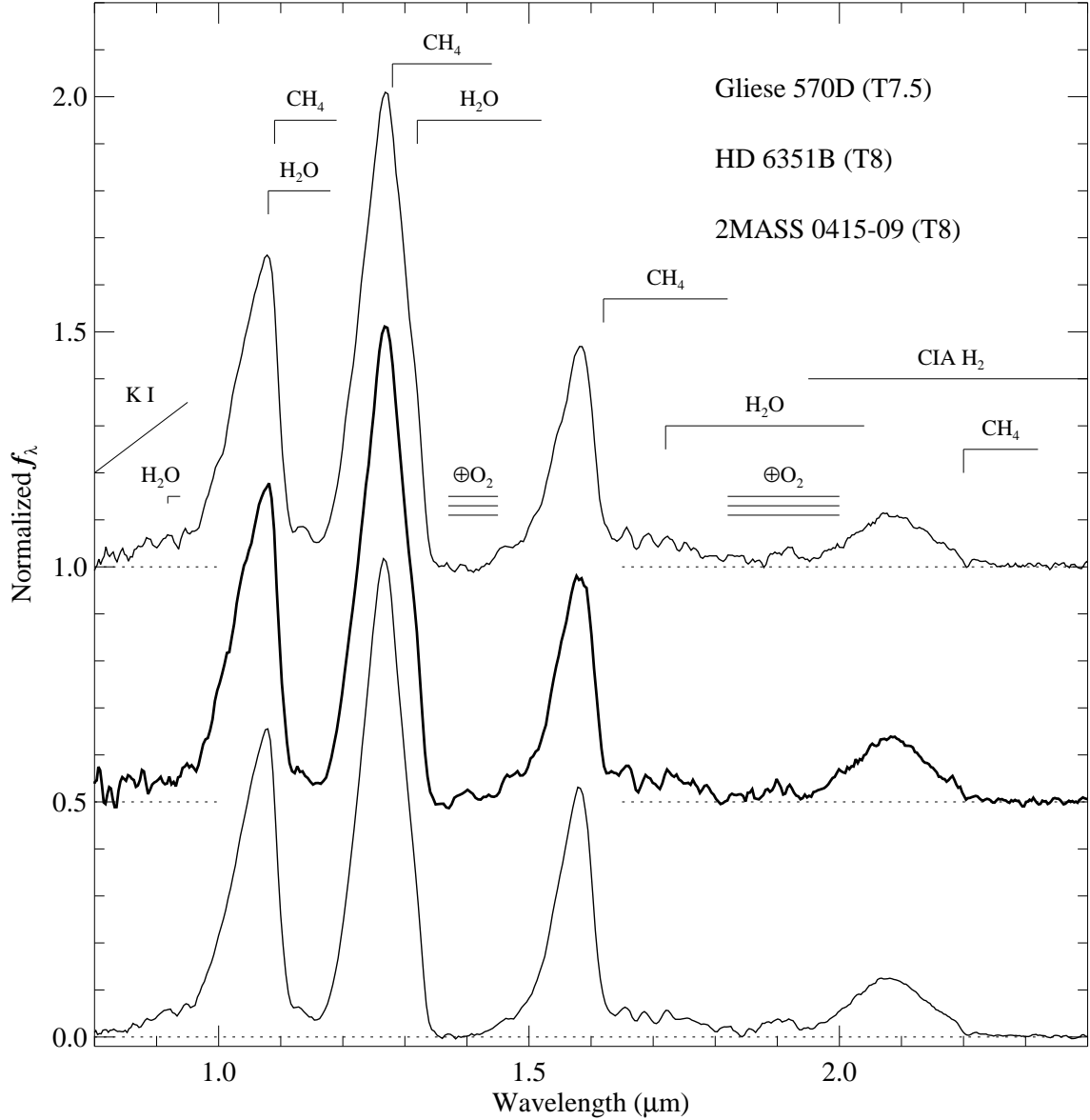}
\caption{From top to bottom, normalized 
near-infrared spectra of the T7.5 companion Gliese~570D,
{\namesrc}B and the T8 field brown dwarf 2MASS~0415-0935, all obtained
with SpeX on IRTF.
All spectra are normalized at 1.27~$\micron$ and offset by
constants (dotted lines).  Major spectral features are noted;
the $\oplus$ symbols designate regions of strong telluric absorption.
\label{fig_spectra}}
\end{figure}

Figure~\ref{fig_spectra} compares the reduced spectrum of {\namesrc}B
to equivalent data for 
the T7.5 companion brown dwarf Gliese~570D \citep{megl570d}
and the T8 field brown dwarf 2MASS~J04151954-0935066 
\citep[hereafter 2MASS~0415-0935]{me02}.
The strong {\water} and {\meth} bands present in the spectrum of {\namesrc}B
are consistent with the observations of \citet{luh06}, and
confirm this object as a late T-type brown dwarf.  
Classification of this source was done using the spectral indices
of \citet{meclass}.  The
measurements and associated subtypes are as follows:
{\water}-J = 0.044 (T8), {\meth}-J = 0.203 (T7.5), {\water}-H = 0.186 (T8), 
{\meth}-H = 0.134 (T7.5) and {\meth}-K = 0.043 (T8), for a mean classification
of T8.  This value is consistent with the T7.5 subtype derived by \citet{luh06} given the typical 0.5 subtype uncertainty in near infrared T dwarf classifications \citep{meclass}.

\section{The Physical Properties of {\namesrc}B}

\subsection{Derived Properties and Metallicity Dependence}

As a companion to a well-studied, nearby star, the properties of 
{\namesrc}B could be readily derived by adopting the age and metallicity of the primary
(e.g.~\citealt{kir01,wil01,luh06}).  Alternately, the 
semi-empirical technique described in BBK can be employed.  
This method involves the
comparison of {\water} and color spectral ratios measured
on the spectrum of a late-type T dwarf, which are separately
sensitive to {\teff} and {\logg} (for a given metallicity),
to the same ratios measured on theoretical spectral models. 
The latter are calibrated to 
reproduce the measured indices for the near infrared SpeX prism
spectrum of Gliese~570D \citep{mewide3}, which has reported physical
parameters of
{\teff} = 782--821~K, 
{\logg} = 4.95--5.23 and [Fe/H] = 0.09$\pm$0.04, based on 
empirical measurements and evolutionary models \citep{geb01,sau06}.   
This method
provides constraints on both {\teff} and {\logg} which can then be used
to infer mass and age with evolutionary models or empirically
via the source's bolometric luminosity.

In BBK and \citet{lie06}, this method was applied assuming solar or subsolar metallicities.  However, the primary of the {\namesrc} has a
metallicity greater than solar, [Fe/H] = 0.12$\pm$0.04 \citep{san04}. 
It has been shown in both BBK and \citet{lie06} that slightly
subsolar metallicities can significantly skew the derived {\teff} and
{\logg} values.  Hence, several cases were considered for
0 $\leq$ [M/H] $\leq$ 0.24.  
The spectral models of \citet{bur06} were employed, spanning
700 $\leq$ {\teff} $\leq$ 1100~K, 4.5 $\leq$ {\logg} $\leq$ 5.5 (cgs)
and 0 $\leq$ [M/H] $\leq$ 0.5.
% interpolated in steps of 
%$\Delta${\teff} = 20~K, $\Delta${\logg} = 0.1 dex and $\Delta$[M/H] = 0.1 dex.
Four pairs of indices were compared
between ({\water}-J, {\water}-H) and (K/J, K/H).\footnote{For 
definitions of these spectral ratios, see BBK and \citet{meclass}.}

\begin{figure}
\plottwo{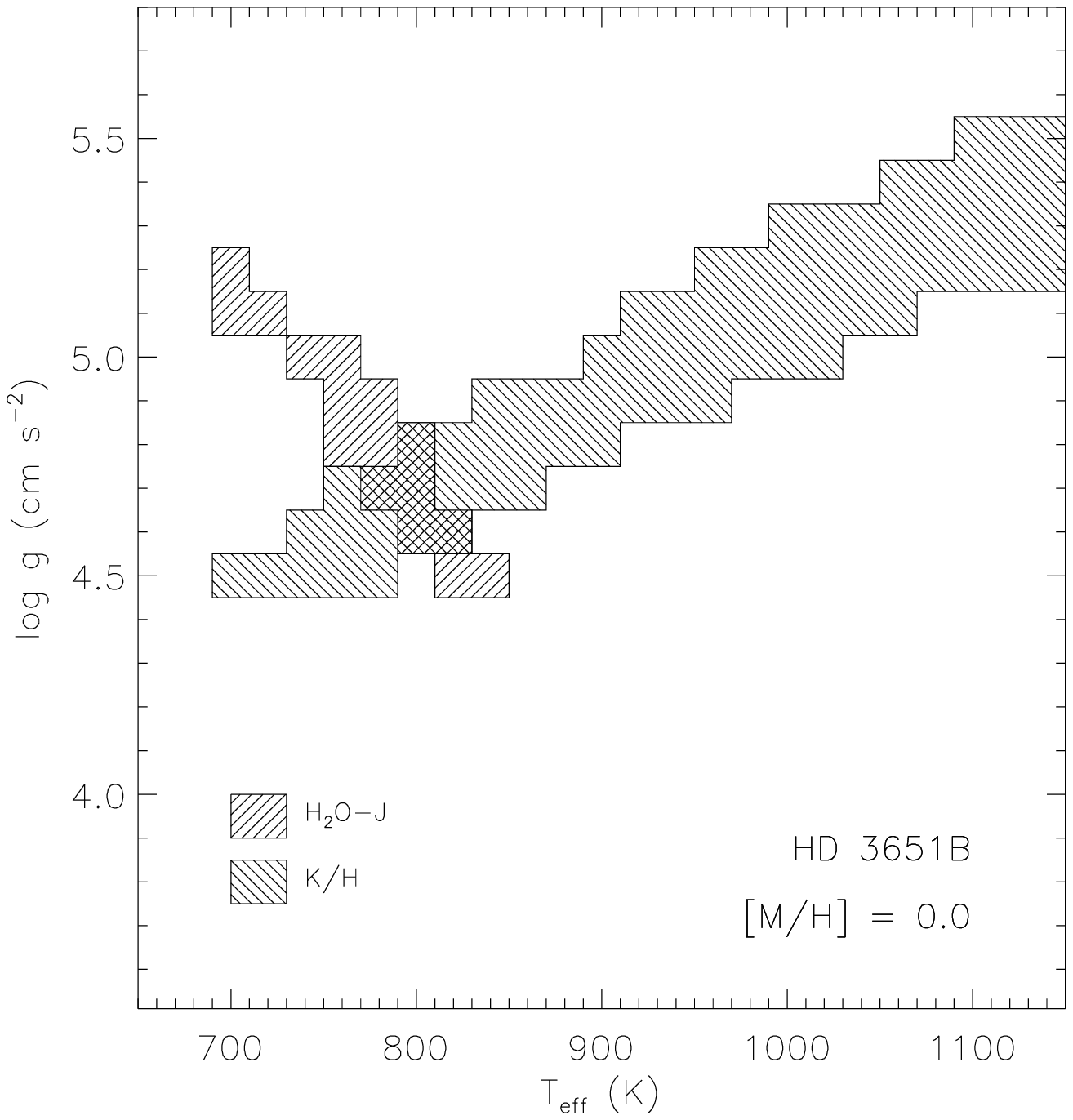}{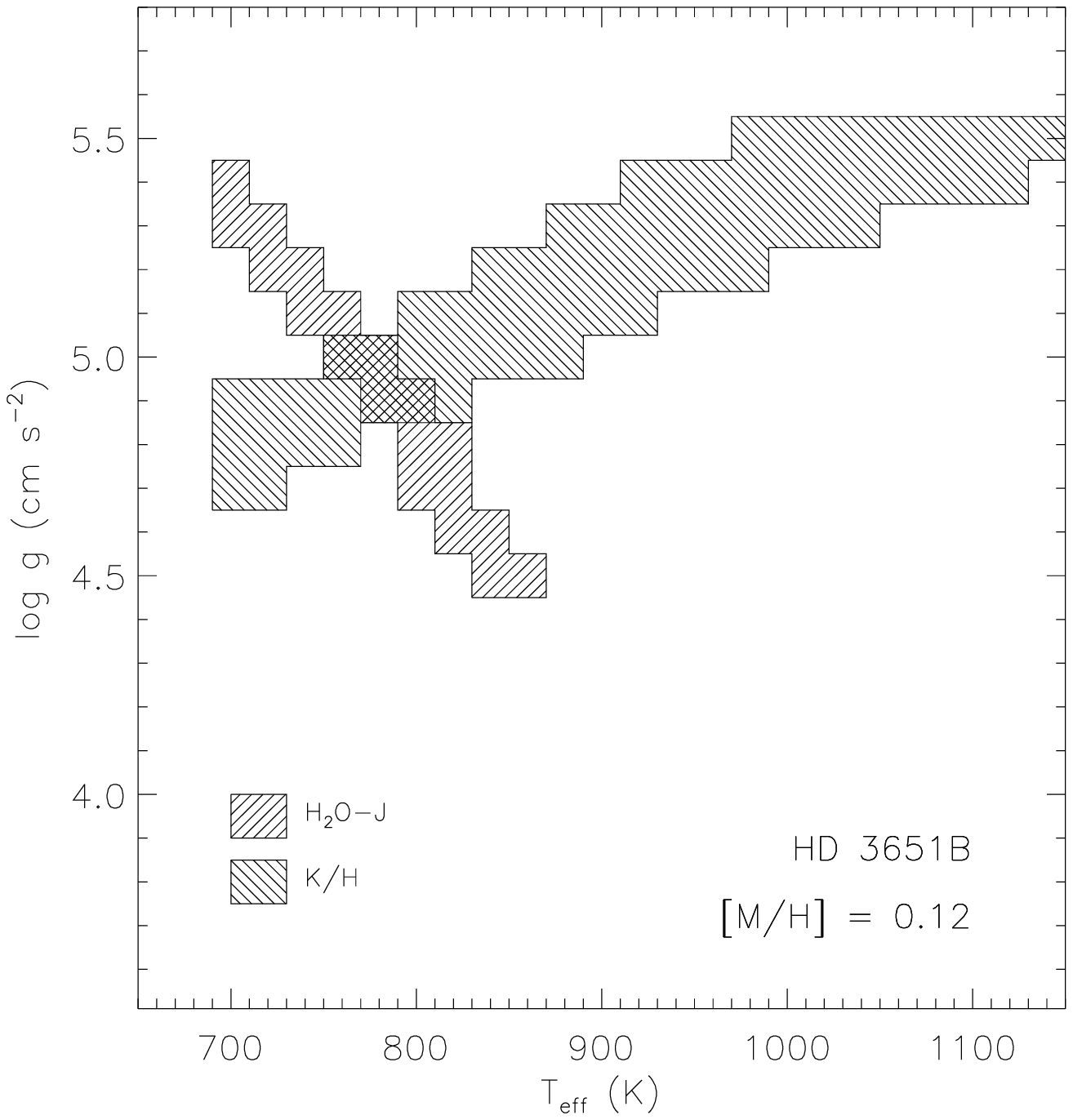}
\caption{Phase spaces of {\teff} and {\logg} constrained 
for {\namesrc}B by
the indices {\water}-J and K/H using the technique of BBK for 
metallicities [M/H] = 0.0 (left) and 0.12 (right). Model indices
are measured from calculations by \citet{bur06} and calibrated
to observations of Gliese~570D.  The intersecting 
cross-hatched regions
provide robust constraints on {\teff} and {\logg}.  
Note the
higher gravity constraint derived for the higher metallicity (see also Table 1).
\label{fig_tg}}
\end{figure}

Figure~\ref{fig_tg} illustrates the ({\teff}, {\logg}) phase
spaces for both [M/H] = 0 and 0.12 over which the spectral ratios 
{\water}-J and K/H measured on the spectrum of {\namesrc}B 
match calibrated values from 
the models. Their intersection yields robust constraints on 
these parameters for the source.  
All four pairs of indices result in a combined 
{\teff} = 790$\pm$30~K
and {\logg} = 5.0$\pm$0.2 
for [M/H] = 0.12 (Table~1); 
including the full uncertainty in [M/H] increases
the uncertainty on {\logg} to 0.3 dex.\footnote{Note that additional systematic
uncertainties of $\pm$50~K and $\pm$0.1~dex in the derived {\teff}
and {\logg} values may
be present due to uncertainty in the parameters of Gliese~570D and 
possible systematic
effects in the spectral models (BBK).}
For [M/H] = 0, the analysis yields {\teff} = 810$\pm$30~K
and {\logg} = 4.7$\pm$0.2.  Note that a decrease in the assumed metallicity leads to a decrease in the derived 
surface gravity.  This is
consistent with trends for subsolar metallicity T dwarfs (BBK; \citealt{lie06}),
and is largely due to the impact of both metallicity and surface gravity
on the relative opacity of collision-induced H$_2$ absorption 
that dominates the $K$-band opacity (e.g., \citealt{sau94}).  

The {\teff} and {\logg} values of {\namesrc}B for [M/H] = 0.12$\pm$0.04
infer M = 0.033$\pm$0.013~M$_{\sun}$
and an age $\tau$ = 0.7--4.7~Gyr, based on the evolutionary models
of \citet{bur01}.  For solar metallicity, these values decrease to
M = 0.020$\pm$0.005~M$_{\sun}$
and $\tau$ = 0.4--1.2~Gyr, primarily due to the lower surface gravity
indicated in this case.  
%As discussed below, this rather young age argues
%against a solar metallicity for this source.
An estimate of mass can also be obtained from the
bolometric luminosity ($L$) of {\namesrc}B by the relation 
M = $Lg/4{\pi}G{\sigma}T_{eff}^4$.
Using the luminosity derived by \citet{luh06}, 
$\log{L/L_{\sun}} = -5.60{\pm}0.05$, and the {\teff} and {\logg} values
derived from the spectroscopic analysis yields 
M = 0.023$_{-0.010}^{+0.020}$~M$_{\sun}$ 
(M = 0.011$_{-0.004}^{+0.008}$~M$_{\sun}$) for [M/H] = 0.12 
([M/H] = 0.0).  These
values are consistent with those derived from the 
evolutionary models, albeit with larger uncertainties 
driven primarily by the uncertainty in {\logg}.

\subsection{HD~3651B: A Metal-Rich T Dwarf}

The method outlined above, based on only two indices, 
cannot discriminate between
[M/H] = 0 and 0.12 for HD 3651B, for in both cases it is possible to derive
robust constraints on both {\teff} and {\logg} (this is not always the case;
cf.\ Figure 8 of BBK).
A third spectral index that is uniquely sensitive to
metallicity is required, and BBK have suggested the ratio between the 
1.05 and 1.25~$\micron$ peak fluxes ($Y/J$ index) as a possible
option.  This ratio is
sensitive to the far red wings of the pressure-broadened
0.77~$\micron$ \ion{K}{1} doublet, and models indicate considerable
metallicity sensitivity to this feature.  However,
the \citet{bur06} models do not reproduce the observed
trends between $Y$-band and $J$-band, likely due to current
uncertainties in the treatment of the far-wing opacities 
\citep{bur03,all03}.  

\begin{figure}
\plotone{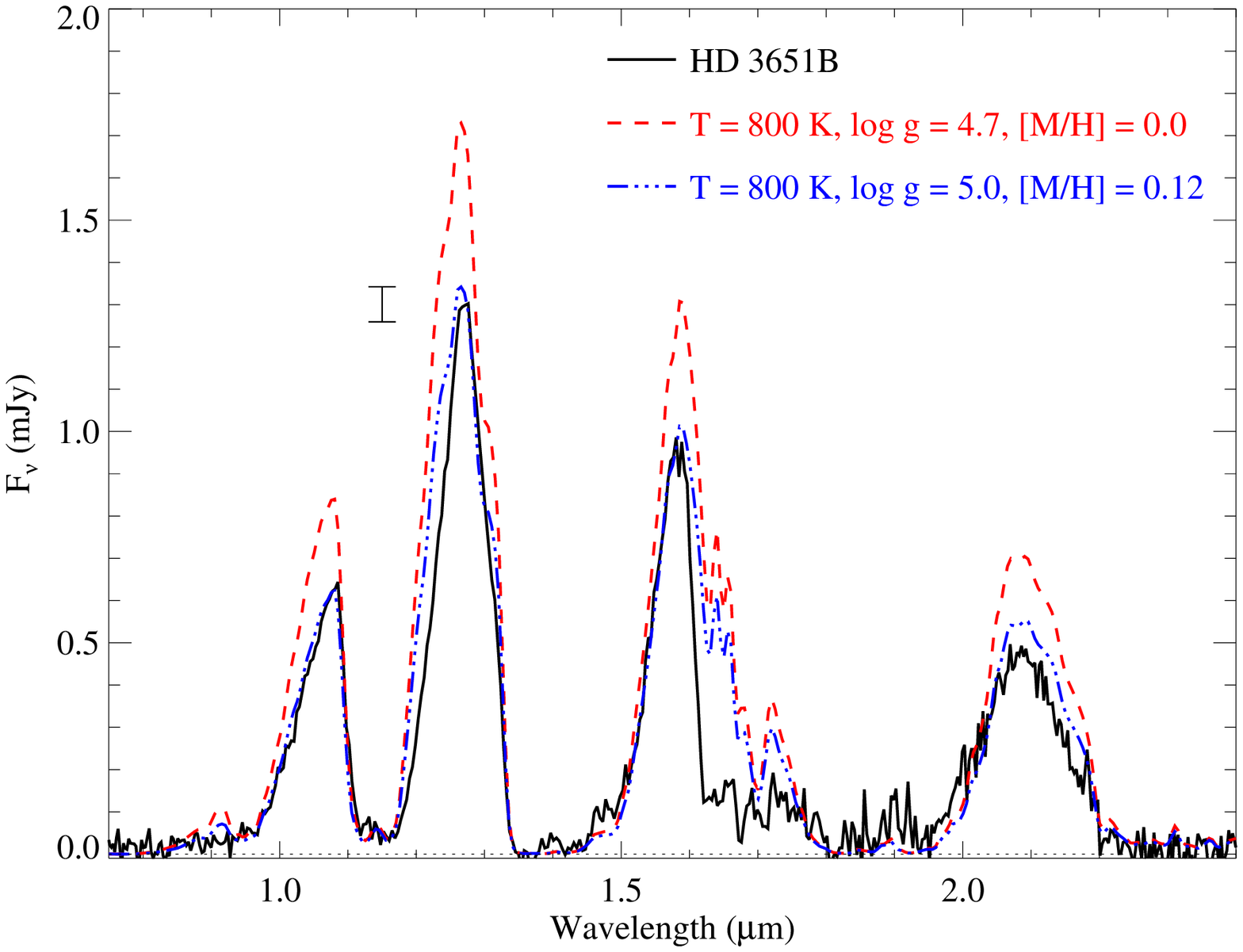}
\caption{Comparison of the absolute flux-calibrated
spectrum of {\namesrc}B (black line) to two {\teff} = 800~K
spectral models from \citet{bur06}: {\logg} = 4.7 and [M/H] = 0.0
(red dashed line), and {\logg} = 5.2 and [M/H] = 0.12 (blue triple-dot
dashed line).  Flux calibration of the {\namesrc}B spectrum is based on 
$J$-band photometry from \citet[16.16$\pm$0.03]{luh06} and the parallax of the primary
as measured by HIPPARCOS \citep[$\pi$ = 90.0$\pm$0.7 mas]{pry97}.
Uncertainty in this calibration at the $J$-band peak is indicated by 
the error bar.
Note the agreement between the data and the [M/H] = 0.12 model.
\label{fig_model}}
\end{figure}

An alternate procedure, used by \citet{lie06},
is to compare the observed absolute fluxes of the brown dwarf
to spectral models based on the best fit parameters for a given metallicity.
This is done in Figure~\ref{fig_model}, which shows the absolute
flux-calibrated spectrum of HD~3651B, based on $J$-band photometry
from \citet[16.16$\pm$0.03]{luh06} and the parallax of HD~3651A from HIPPARCOS \citep[$\pi$ = 90.0$\pm$0.7 mas]{pry97};
to predicted fluxes for two {\teff} = 800~K models, 
one with {\logg} = 4.7 and [M/H] = 0 and
one with {\logg} = 5.0 and [M/H] = 0.12.  While the models
provide poor fits to the 1.1 and 1.6~$\micron$ {\meth} bands, the result
of well-documented deficiencies in {\meth} opacities at these
wavelengths, it is
clear that the lower
surface gravity, solar metallicity model is too bright as compared
to the data.  The [M/H] = 0.12 model, on the other hand, provides a 
reasonably good match to the absolute fluxes at each of 
the spectral peaks.  This comparison clearly supports a supersolar
metallicity for HD~3651B.

Another consideration is the inferred age of {\namesrc}B as 
compared to previous age determinations for
{\namesrc}A.   Published estimates for the primary 
vary appreciably.
\citet{wri04} report an age of 5.9~Gyr on the basis of weak
\ion{Ca}{2} HK emission ($\log{R_{HK}^{\prime}} = -5.07$) and the age/activity relation of \citet{don93}.  On the other hand,
\citet{roc04} report a chromospheric age of only 2.1~Gyr based on a
higher value of $\log{R_{HK}^{\prime}} = -4.85$ from
\citet{sod93}. Indeed, a 35\% variation in \ion{Ca}{2} 
HK emission of {\namesrc}B
has been seen over 17 years of observations at Mount 
Wilson Observatory \citep{dun91}.
From isochrone fitting, \citet{nor04} derive a minimum age
of 2.6~Gyr, \citet{val05} an age of 8$^{+4}_{-5}$~Gyr and 
\citet{tak06} a minimum age of 11.8~Gyr.  
While there appears to be little consensus,
an older age for {\namesrc}A is most consistent with its
chromospheric inactivity,
its long rotation period
(48~days; \citealt{noy84})
and its $UVW$ space velocities \citep{egg89}.  A value in the range
2--12~Gyr is indicated.  

This result again argues in favor of 
HD~3651B being metal-rich.  
The derived age for the solar metallicity case is considerably younger
than that of the primary; while
for [M/H] = 0.12$\pm$0.04, the ages of the primary
and secondary overlap
in the range of 2--5~Gyr.  Note that this age constraint 
is more precise than that inferred from the primary alone, and is 
consistent with both components having been formed at the same
time and from the same gas/dust reservior (same age and metallicity)

There are two conclusions that may be drawn from this
analysis. First, it is clear that the
determination of {\teff} and {\logg}, and hence mass and age,
of an individual brown dwarf using the BBK technique is highly
sensitive to the assumed metallicity, even for small variations.\footnote{[M/H] = 0.12 implies only a 32\% increase in metal abundance over solar.}
This reflects the overall sensitivity of cool brown dwarf spectra to changes
in photospheric abundances, and is not surprising given that the 
entire near infrared spectrum of a T dwarf is draped by atomic and molecular 
absorptions.  While metallicity effects have been previously noted 
in the spectra of metal-poor
T dwarfs such as 2MASS~J09373487+2931409 \citep[BBK]{me02,kna04}
and 2MASS~J12373919+6526148 \citep{me99,lie06}, it is now clear 
that metal-rich 
T dwarf spectra can also exhibit metallicity effects.  These results emphasize
the importance of characterizing the overall abundances --- and abundance
patterns --- in deriving a complete and accurate characterization
of a low-temperature brown dwarf.  
Note that HD~3651B and Gliese~570D may be a particularly useful
pair in studying metallicity effects, as they appear to be identical in nearly all aspects ({\teff}, {\logg}, age, mass and spectral energy distribution)
except metallicity.

Second, it is possible that coeval 
brown dwarf companions may provide
additional or even better constraints on the ages of their host systems than some 
stellar primaries.   While age estimates for stars 
on the basis of chromospheric activity and kinematics can vary
appreciably, and are less discriminating at later ages, the example of HD~3651B demonstrates that a late-type brown dwarf can potentially be age-dated with higher precision.  However, the accuracy of these
ages remains unclear, as they are tied to both 
the fidelity of evolutionary
models and, in the case of the BBK technique, the accuracy and calibration of 
spectral models.  
In the former case, there have been few empirical checks on the evolutionary
models of brown dwarfs at late ages,\footnote{At very young ages ($\sim$1--5~Myr), both
\citet{moh04} and \citet{sta06} have found discrepancies between
evolutionary models and observed parameters.  It is possible that these discrepancies arise from the initial conditions assumed in the evolutionary
models, and may be unimportant at late ages (cf. \citealt{mar06}).} although BBK found consistency between the 
radii of field T dwarfs empirically derived from luminosity measurements and those
based on evolutionary models.
In the latter case, there appears to be consistency between
different sets of spectral models, as an equivalent
analysis of the spectrum of {\namesrc}B 
using the COND models of \citet{all01} produced nearly identical
results.  Hence, at first glance it appears that 
late-type brown 
dwarf companions could be useful sources for characterizing the properties
of a presumably coeval system, although further empirical
tests of brown dwarf evolutionary models
are needed.

\section{{\namesrc}B: An Extrasolar Nemesis?}

{\namesrc}B is the first wide brown dwarf companion
found in a system
hosting at least one known planet.  This raises an intriguing question: 
how has the evolution of the {\namesrc} planetary system ---
and specifically the development
of life on any putative habitable planets in this system --- been influenced by 
this distant brown dwarf companion?   
This speculative question is an appropriate one, 
as putative terrestrial bodies within a portion of
{\namesrc}A's habitable zone \citep{kas93}
can persist without dynamical disruption by
{\namesrc}b \citep{jon05}, even considering this giant planet's 
probable inward migration \citep{man03}.
Terrestrial planet formation around the primary of a
wide binary system has also been shown to be feasible \citep{hol97,qui04}.
However, the formation, existence and habitability of terrestrial planets
in the {\namesrc} system may be inhibited by 
dynamical perturbations from {\namesrc}B, both directly through orbital
modulation\footnote{Indeed, \citealt{luh06} suggest that the eccentricity of {\namesrc}b
may be driven by secular perturbations from {\namesrc}B.} \citep{tak05}; or the scattering of small bodies, resulting in sustained impacting onto the planet's
surface.  Yet terrestrial planets on
mildly eccentric orbits can maintain habitability depending on the properties
of their atmospheres \citep{ada06}, while increased impacting rates 
may in fact facilitate the emergence of life. 
In the case of the 
Earth, the late heavy bombardment period between 4.5 and 3.8 Gyr ago
resulted in one particularly cataclysmic impact that
formed the Moon \citep{har75}, 
whose tidal forces have maintained Earth's tilt to the Sun and have
thereby reduced climate variation.   Impacts of icy bodies 
may have brought necessary water ice 
to Earth's surface \citep{chy87,del98} as well as chemical
precursors to biotic life \citep{chy90}.  On the other hand,
later impacts clearly have a negative effect on the evolution
of macrobiotic life \citep{alv80,hil91}.  

As an instigator of 
planetary impacts, {\namesrc}B may play the role of Nemesis in
the {\namesrc} system.  This
hypothesized companion to the Sun \citep{dav84,whi84}
has been proposed to
explain apparent periodicities in massive extinctions 
\citep{fis77,rau84,hut87},
terrestrial cratering rates \citep{alv84,ram84}, reversals
of Earth's magnetic fields \citep{rau85} and, more recently,
the peculiar orbits of inner Oort Cloud 
planetoids \citep{bro04,gau05}.  If  
{\namesrc}B were present at its observed separation 
but bound to the Sun, 
it would be invisible to the naked eye ($M_V \sim 27$ implies $V \sim 9$)
but readily detectable by Hipparcos, Tycho \citep{pry97}
and early wide-field near-infrared sky surveys ($K \sim -1$; cf. the
Two-Micron Sky Survey, \citealt{neu69}).  If HD~3651B
was at the distance from the Sun proposed for Nemesis 
($\sim$1.5$\times$10$^5$~AU; \citealt{dav84}), it
would be invisible on visual photographic plate images 
($V$ $\sim$ 20)
but a relatively bright source ($J$ $\sim$ 10)
in the Two Micron All Sky Survey \citep{skr06}.

Searches for a solar Nemesis have so far turned up negative
(e.g., \citealt{per86}).  However, as widely-separated low mass
stellar companions appear to be
prevalent in known planet-hosting systems (e.g., \citealt{rag06}),
and with widely-separated 
brown dwarf companions to these systems only now being identified,
Nemesis counterparts may be a common facet of all planetary systems
and an important consideration for the emergence of terrestrial-based life.

\acknowledgments

The author thanks D.\ Chakrabarty and J.\ Winn
for their helpful comments on an early version of the manuscript,
and acknowledges the support of telescope operator E.\ Volquardsen
and instrument specialist J.\ Rayner during the IRTF observations.
The author also thanks the anonymous referee for her/his insightful comments
that led to the investigation of metallicity effects in the {\teff}/{\logg}
analysis.  
Electronic versions of their evolutionary and spectral models 
were kindly provided by A.\ Burrows and P.\ Hauschildt for this work. 
The author wishes to recognize and acknowledge the 
very significant cultural role and reverence that 
the summit of Mauna Kea has always had within the 
indigenous Hawaiian community.  He is most fortunate 
to have had the opportunity to conduct observations from this mountain.

Facilities: \facility{IRTF(SpeX)}

\clearpage

\begin{deluxetable}{lccccc}
\tabletypesize{\small}
\tablecaption{Derived Physical Properties of {\namesrc}B.}
\tablewidth{0pt}
\tablehead{
\colhead{Parameter} &
\colhead{[M/H] = 0.00} &
\colhead{[M/H] = 0.08} &
\colhead{[M/H] = 0.12\tablenotemark{a}} &
\colhead{[M/H] = 0.16} &
\colhead{[M/H] = 0.24} \\
}
\startdata
{\teff}~(K) &  810$\pm$30  & 790$\pm$30 & 790$\pm$30 & 790$\pm$30 & 780$\pm$40 \\
{\logg}~(cgs) &  4.7$\pm$0.2 & 4.9$\pm$0.2 & 5.0$\pm$0.2  & 5.0$\pm$0.2 & 5.2$\pm$0.2 \\
Mass~(M$_{\sun}$) & 0.020$\pm$0.005 & 0.027$\pm$0.007 & 0.032$\pm$0.008 & 0.035$\pm$0.011 & 0.047$\pm$0.013 \\
Age (Gyr) & 0.4--1.2 &  0.7--2.8 & 1.0--3.7 & 1.0--4.7 & 2.6--9.3 \\
\enddata
\tablenotetext{a}{\citet{san04} derive a metallicity for the primary of [Fe/H] = 0.12$\pm$0.04.}
\end{deluxetable}

\clearpage

\clearpage

\end{document}